# Spin wave assisted current induced magnetic domain wall motion

Mahdi Jamali,[1] Hyunsoo Yang,[1,a)] and Kyung-Jin Lee [2]

[1]Department of Electrical and Computer Engineering, National University of Singapore, Singapore

[2]Department of Materials Science and Engineering, Korea University, Seoul, Korea

The interaction between the propagating spin waves and the current driven motion of a transverse domain wall in magnetic nanowires is studied by micromagnetic simulations. If the speed of domain walls due to current induced spin transfer torque is comparable to the velocity driven by spin waves, the speed of domain wall is improved by applying spin waves. The domain wall velocity can be manipulated by the frequency and amplitude of spin waves. The effect of spin waves is suppressed in the high current density regime in which the domain wall is mostly driven by current induced spin transfer torque.

a) e-mail address: eleyang@nus.edu.sg

Current induced domain wall motion in patterned magnetic nanowires has generated wide interest among researchers due to its potential applications in magnetic memories [1-3] and logic devices [4-5]. The motion of domain walls due to spin transfer torque (STT) of electrons has been well studied theoretically [6-8] and experimentally [9-14]. Recently a few studies have been reported on the interaction of spin waves and domain walls [15-16], domain wall induced phase shifts in spin waves [17], and spin wave generation by domain walls [18-19]. It has been also shown that propagating spin waves can move domain walls [20] and the STT effect and spin waves can interact with each other [21]. However, the interaction of spin waves and domain walls in the presence of current induced STT has not been reported.

In this letter, we report the interaction of propagating spin waves and domain walls in magnetic nanowires in the presence of electrical current. The present study aims to understand the effect of spin waves with different frequencies on the motion of domain walls in both low and high current density. It is found that propagating spin waves can enhance the domain wall velocity when the velocity of domain wall caused by STT is comparable to the domain wall velocity when driven by spin waves. Moreover, depending on the excitation frequency of spin waves, the domain wall motion is controllable in the low current density regime.

The structure that we have used in our simulations is shown in Fig. 1(a). The wire has 3005 nm length, 50 nm width, and 10 nm thickness and a transverse domain wall was placed at the center of the wire. The simulation cell size is 5×5×5 nm$^3$ and the nanowire is made of Permalloy (Py) with the saturation magnetization ($M_S$) of 860×10$^3$ A/m, the exchange stiffness ($A_{ex}$) of 1.3×10$^{-11}$ J/m, and the Gilbert damping constant ($\alpha$) of 0.01.

We used the object oriented micromagnetic framework (OOMMF) [22] code for simulations that solves the Landau-Lifshitz-Gilbert (LLG) equation which incorporates the spin transfer torque term for the domain wall motion [23].

In order to generate spin waves, we applied an external magnetic field to the first 5 ×50×10 nm$^3$ in the left side of the structure. The applied field varied sinusoidally in time with frequency ($f$) and amplitude $H_0$ as $H = H_0 \sin(2\pi t f) \hat{y}$ in the $y$-direction. In order to apply current induced STT, we have injected current in the negative $x$-direction where electrons flow in positive $x$-direction.

The LLG equation including the spin torques can be written as [19]:

$$\frac{\partial M}{\partial t} = -\gamma_0 H_{eff} \times M + \frac{\alpha}{M_s} M \times \frac{\partial M}{\partial t} + T_a + T_{na} \quad (1)$$

where $\gamma_0$ is the gyromagnetic constant. In this equation $T_a$ and $T_{na}$ are adiabatic and nonadiabatic torque terms, respectively, and they are defined as follows:

$$T_a = -(u.\nabla)M \quad (2)$$

$$T_{na} = \frac{\beta}{M_s} M \times [(u.\nabla)M] \quad (3)$$

The dimensionless coefficient $\beta$ characterizes the nonadiabatic contribution which is 0.1 in our simulations and the $u$ parameter is the effective drift velocity of the conduction electron spins defined by $u = JPg\mu_B/(2eM_S)$, where $J$ is the current density, $P$ is the spin polarization, $\mu_B$ is the Bohr magneton, and $e$ is the electron charge. If we assume $P = 0.7$ and $u = 5$ m/s, the current density is $J=1.1\times10^{11}$ A/m$^2$. In order to prevent the reflection of spin waves from the magnetic nanowire edges, we applied anti-reflection boundary conditions [21] by increasing the damping coefficient near the nanowire edges to unity.

Figure 1(b) shows that a domain wall is displaced by exciting only spin waves with a frequency of 18 GHz and different magnetic field amplitudes. The excitation frequency of 18 GHz is chosen based on the previous observation where the same size of the nanowire is used [20] and it is about twice of the resonance frequency of the domain wall (8.9 GHz). It is clear that domain wall velocity increases with increasing the magnetic field amplitude. The average domain wall velocity has been calculated based on time required to displace a domain wall between 5 nm and 150 nm from its initial position. As seen in Fig. 1(c), the average domain wall velocity is monotonously proportional with increasing the magnetic field amplitude similar to the previous report [20].

In order to see the interaction between spin waves and a domain wall in the presence of current, we injected different current densities into the nanowire and calculated the domain wall velocity by taking into account the effect of STT. Here we fixed the spin wave frequency at 18 GHz. In the low current density ($u < 20$ m/s) case, spin waves can enhance the speed of domain wall as shown in Fig. 1(d). For example, when $u = 5$ m/s, the domain wall velocity is 54.6 m/s with a field amplitude of 2 kOe and the domain wall velocity increases to 81.6 m/s for a field amplitude of 10 kOe. For $u$ greater than 20 m/s the domain wall velocity is independent of the excitation field amplitude of spin waves. The domain wall velocity increases linearly with increasing the current density regardless of the field amplitude. In our simulation, spin waves and current are simultaneously injected. In order to test whether the delay between spin waves and current make any change in the total velocity of domain wall, spin waves is turned on first and current is injected after 5 ns. As shown in Fig. 1(d), there is no effect on the domain wall velocity.

In Fig. 2(a) the domain wall displacement is shown as a function of time with $u =$ 5 m/s and $f =$ 18 GHz for the various field amplitudes. By increasing the field amplitude, domain wall displacement increases after 2 ns which is required for the domain wall start to move. In Fig. 2(b) the average velocity of domain walls is shown for $u =$ 5 m/s and $f =$ 18 GHz. The domain wall velocity increases with increasing the excitation field amplitude and the relation is mostly linear. When we increase the current density to $u =$ 50 m/s, the time evolution of domain walls is almost the same for all fields as shown in Fig. 2(c). In the high current density regime the domain wall speed is independent of the spin wave amplitude as shown in Fig. 2 (d). We found that the critical current, above which the contribution of spin waves on the domain wall velocity is less than 1%, therefore, we can neglect the spin wave effect on the domain wall motion, is proportional to the excitation field. For example, when $H_0 =$ 2, 5, and 10 kOe, the critical current $u$ is 8, 15 and 20 m/s, respectively. The reason for the transition between these two regimes requires further studies for better understanding.

It is interesting to note that the spin wave velocity is much faster than that of the domain wall motion velocity. The wavelength of spin wave is around 60 nm in 18 GHz excitation frequency which results in a phase velocity ($v = \omega / k$) of 1080 m/s. Although the speed of spin waves is higher than the speed of domain wall due to STT, spin waves need interaction time with domain walls, therefore, the speed of domain wall due to spin wave alone (~ 45 m/s) is much less than speed of spin wave itself (1080 m/s).

We believe that the gradient of spin waves across the domain wall could be one of the driving forces of domain wall motion. When spin waves interact with a domain wall, the magnetization inside a domain wall starts to oscillate with a resonance frequency of the domain wall (8.9 GHz) and then the oscillation frequency changes to the spin wave excitation frequency. We still need more studies to fully understand the detail mechanism.

In order to investigate the effect of the frequency of spin waves on the domain wall velocity, we applied spin waves with different frequencies. As shown in Fig. 3(a), the domain wall displacement varies by changing the excitation frequency. In Fig. 3(b) the velocity of the domain wall is shown against the excitation frequencies for $u = 5$ m/s and $H_0 = 10$ kOe. It can be seen that in a low current density ($u = 5$ m/s), spin wave will enhance the speed of domain wall. If the spin wave excitation frequency is less than the cut-off frequency of the nanowire, spin waves cannot propagate through the nanowire, therefore the velocity of domain wall becomes constant for less than 12 GHz as shown in Fig. 3(b) and only current induced STT contributes to the domain wall motion. By using the approximate dispersion relation [24], we obtained a cut-off frequency of 12.9 GHz. In order to find the cut-off frequency of the nanowire through simulation, we applied a Gaussian pulse and measured the frequency response of the nanowire at different points [24-26]. Here we used the pulse width of 20 ps and the amplitude of 100 Oe and then by using Fast Fourier Transformation (FFT), we calculated the frequency of spin wave that can propagate along the nanowire. It is clear from Fig. 3(e) that spin waves with the frequency less than 12 GHz will be mostly attenuated which shows a good agreement with the approximated dispersion relation.

We also studied the high current regime by increasing *u* to 50 m/s. The domain wall displacement is almost independent of the spin wave frequency as shown in Fig. 3(c). If we compare the domain wall velocity at different frequencies as summarized in Fig. 3(d), we observe that the change in the speed of domain wall due to spin waves is less than 1% even by changing the frequency of spin waves up to 32 GHz confirming that the relative velocity of domain wall due to spin waves and current induced STT is an important parameter in order to observe the effect of spin waves on the current induced domain wall motion.

In summary, we have shown that spin waves can enhance the current induced domain wall velocity in the low current regime where the domain wall velocity due to current induced STT is comparable to the domain wall velocity when driven by spin waves. We can manipulate this enhancement by changing the excitation amplitude and frequency of spin waves in the same regime. Our results demonstrate that spin waves interaction with current induced STT should be carefully treated in the domain wall motion studies and provide a way to control the domain wall motion for magnetic nano-devices.

Figure Captions

Figure 1. (a) Schematic illustration of a magnetic nanowire with a transverse domain wall at 1505 nm from the left edge of the nanowire. (b) Domain wall displacements due to spin waves with different field amplitudes. (c) The domain wall velocity versus magnetic field amplitude of spin waves. (d) The domain wall velocity at different current densities with different field amplitudes.

Figure 2. (a) Domain wall displacements versus time for $u = 5$ m/s and $f = 18$ GHz. (b) The domain wall velocity for $u = 5$ m/s and $f = 18$ GHz with different excitation amplitudes. (c) Domain wall displacements versus time for $u = 50$ m/s and $f = 18$ GHz. (d) The domain wall velocity for $u = 50$ m/s and $f = 18$ GHz with different excitation field amplitudes.

Figure 3. (a) Domain wall displacements for $u = 5$ m/s and $H_0 = 10$ kOe with different frequencies. (b) The domain wall velocity versus frequency for $u = 5$ m/s and $H_0 = 10$ kOe. (c) Domain wall displacements for $u = 50$ m/s and $H_0 = 10$ kOe with different frequencies. (d) The domain wall velocity versus frequency for $u = 50$ m/s and $H_0 = 10$ kOe. (e) Frequency spectral image along the $x$-axis of the $M_z$ component for Gaussian pulse excitation.

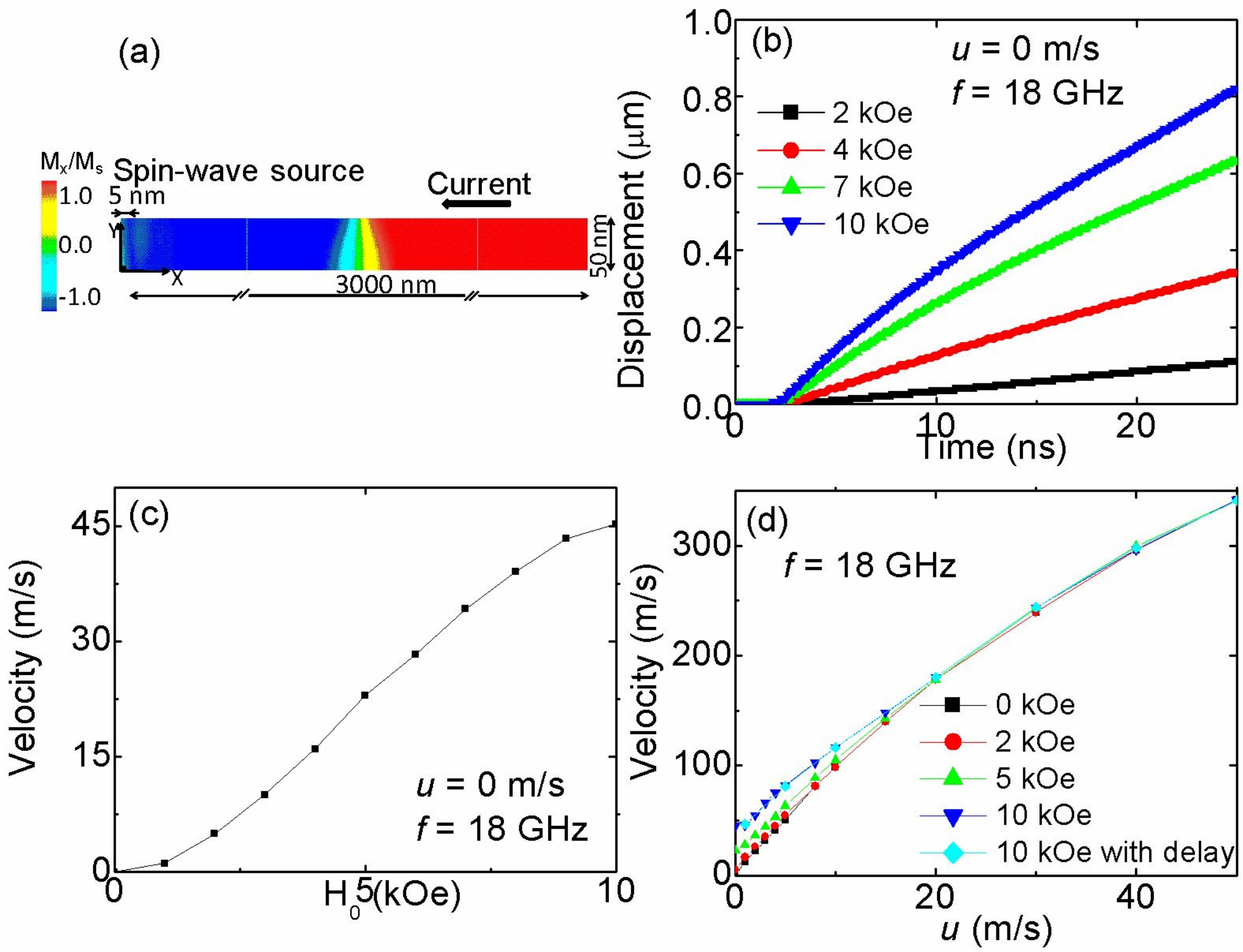

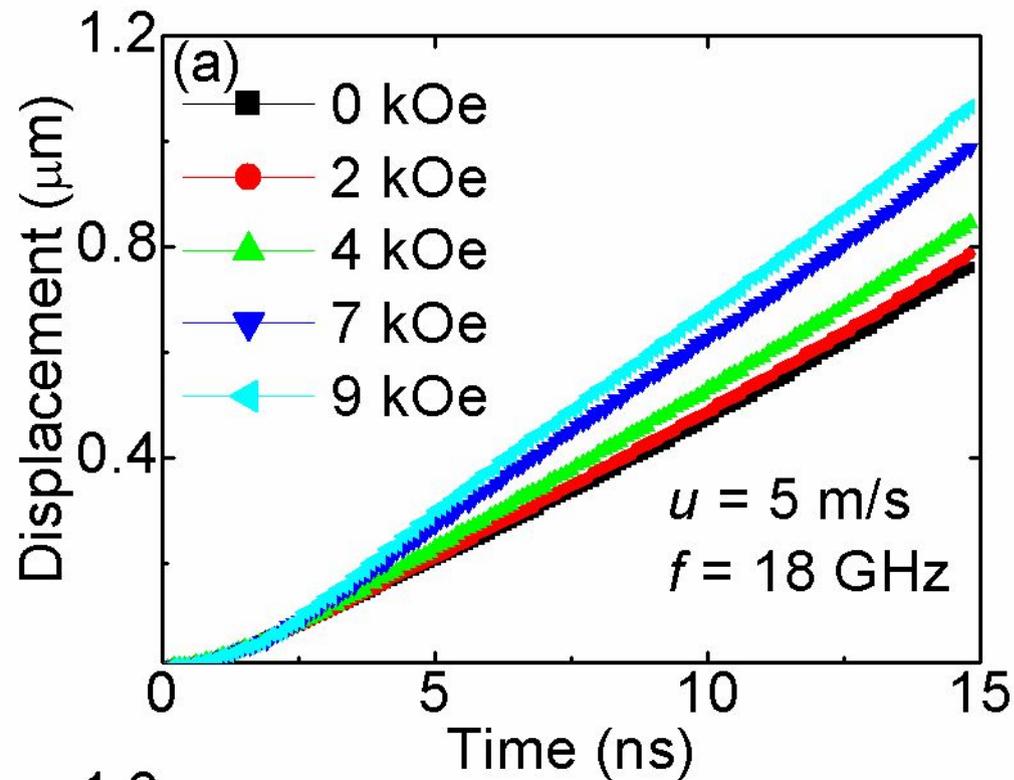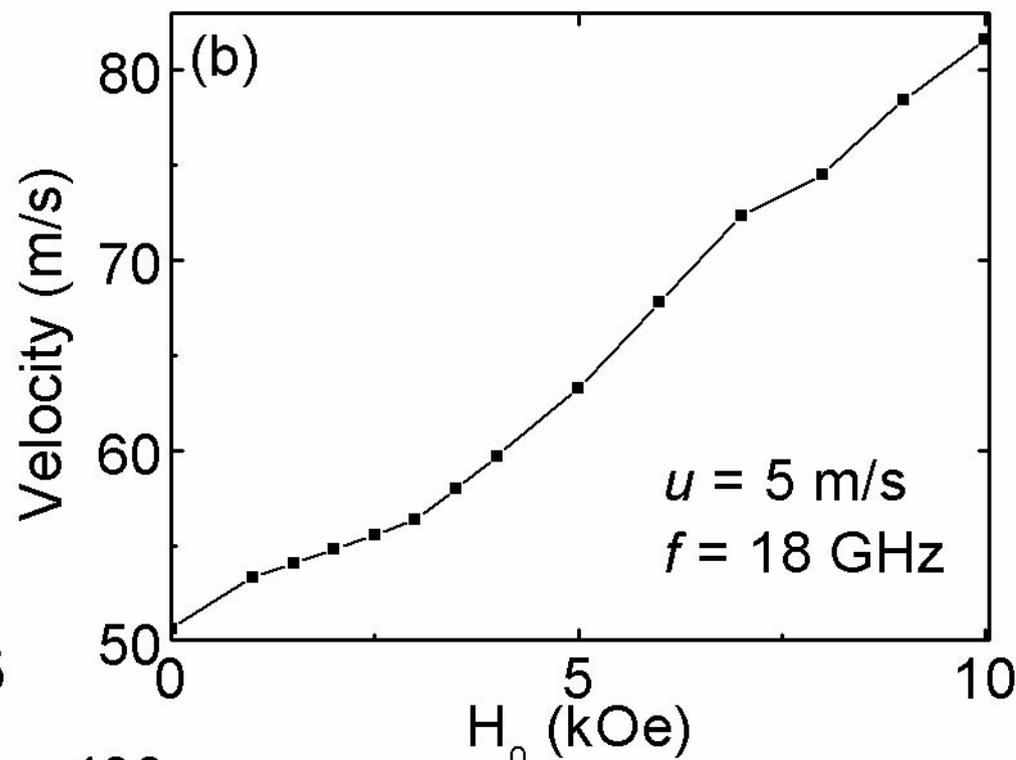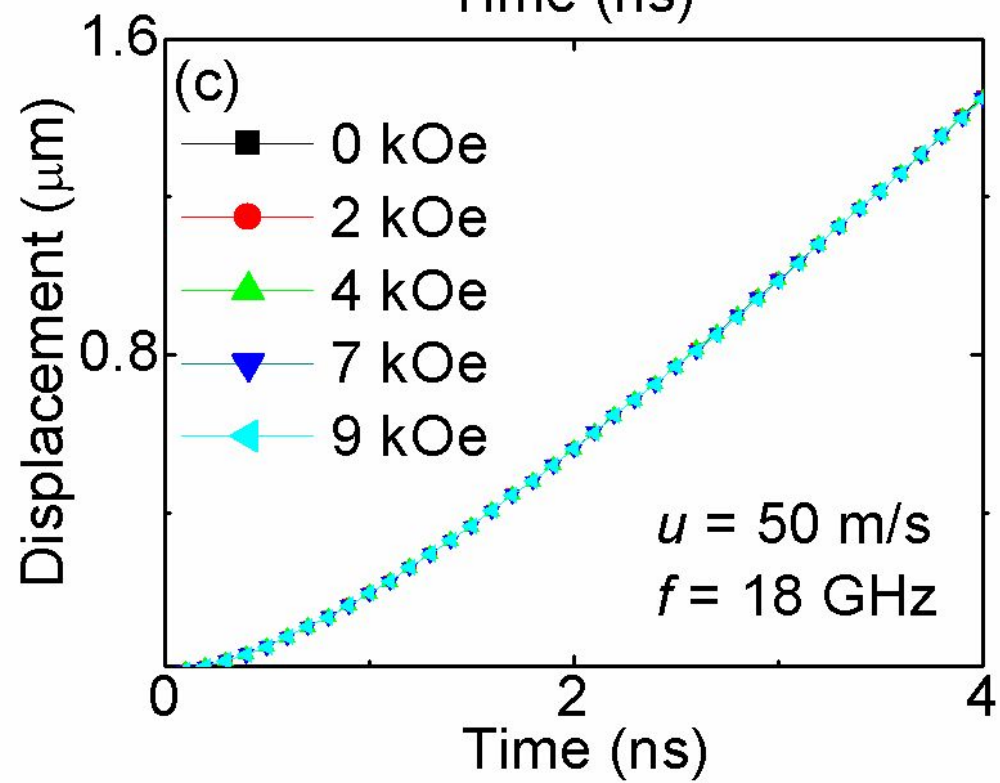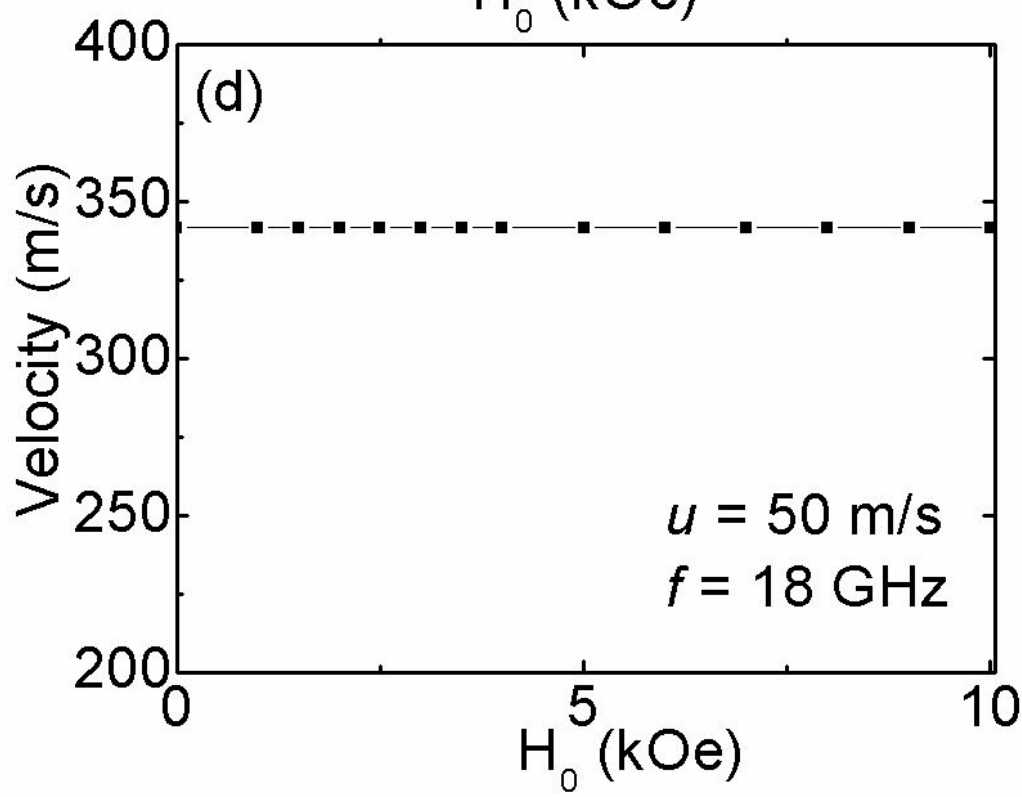

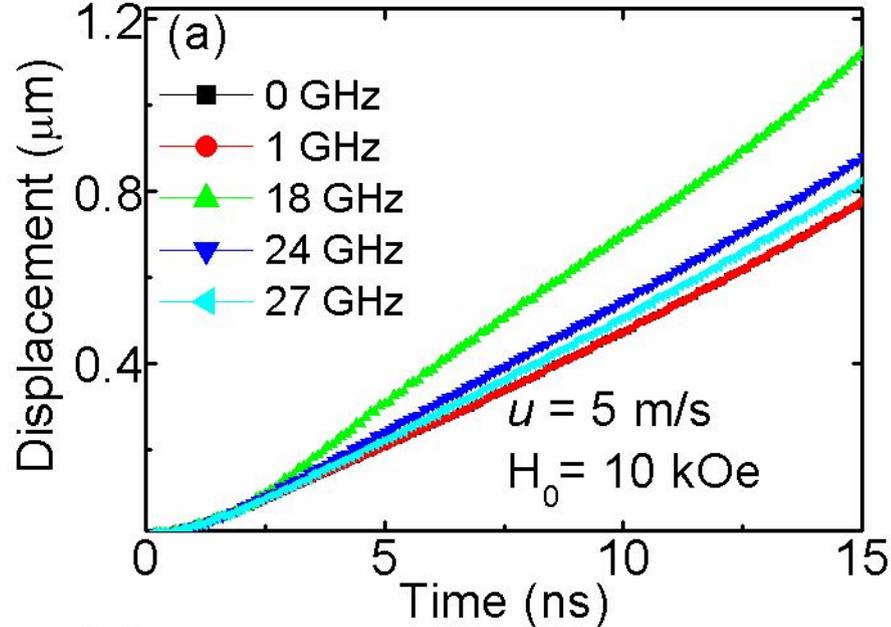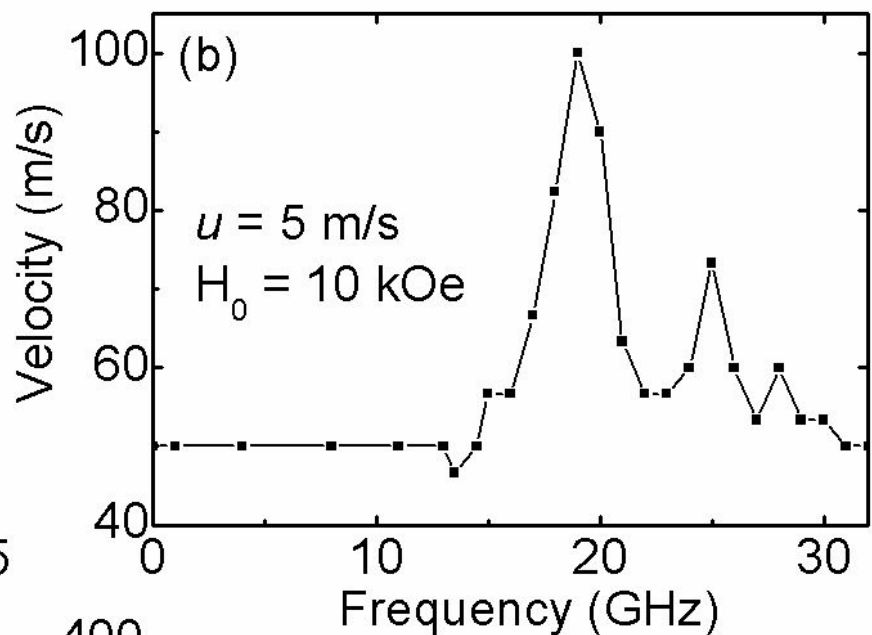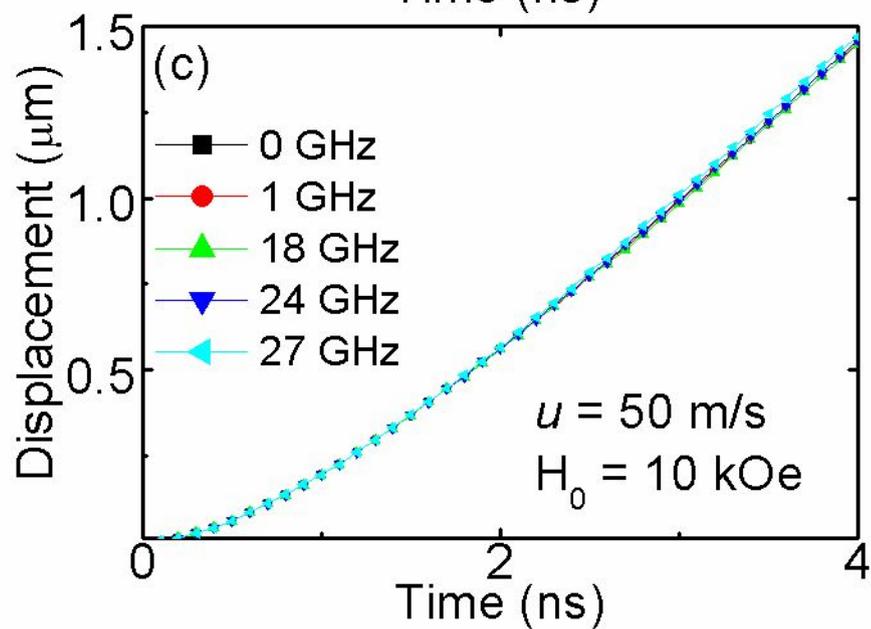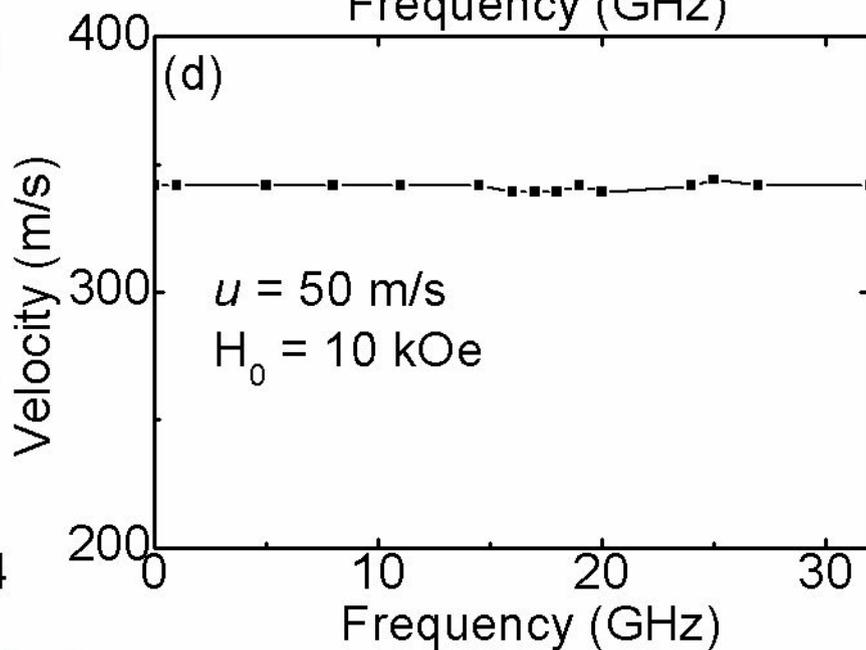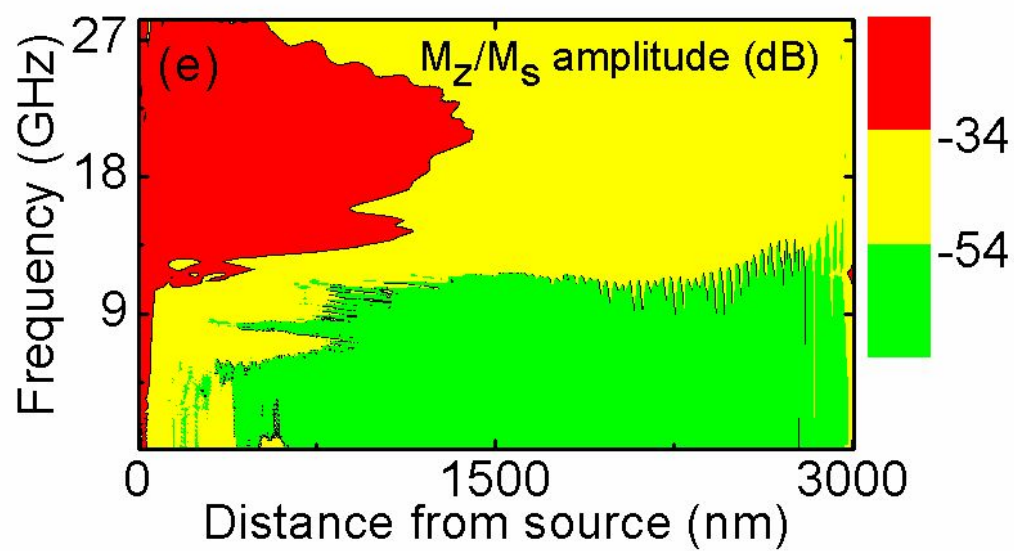